\documentclass[aps,twocolumn,groupedaddress,floatfix]{revtex4}

\usepackage{epsfig}

\begin{document}

\title{Semiclassical theory of energy diffusive escape}

\author{Alvise Verso and Joachim Ankerhold}

\affiliation{Institut f\"ur Theoretische Physik, Universit\"at Ulm, Albert-Einstein-Allee 11,
89069 Ulm, Germany}

\date{\today}

\begin{abstract}
Thermal escape out of a metastable well is considered in the weak friction regime, where the bottleneck for decay is energy diffusion, and at lower temperatures, where quantum tunneling becomes relevant. Within a systematic semiclassical formalism an extension of the classical diffusion equation is derived starting from a quantum mechanical master equation. In contrast to previous approaches finite barrier transmission also affects transition probabilities. The decay rate is obtained from the stationary non-equilibrium solution and captures the intimate interplay between thermal and quantum fluctuations above the crossover to the deep quantum regime.
\end{abstract}

\maketitle

\section{Introduction}
Thermally activated escape over high energy barriers can be found in a huge variety of physical and chemical
processes \cite{benson_60,haenggi_90,haenggi_93,pollak_05}.  While these issues have been studied for quite a while now, the subject has gained new interest in the context of quantum information processing, where read-out devices for superconducting quantum bits based on Josephson junctions (JJs) have been implemented \cite{vion_02,martinis_02,claudon_04}. In these realizations the exponential sensitivity of the switching rate out of the zero voltage state on the barrier topology has been exploited to discriminate the qubit states. Recently, a new type of detector (Josephson Bifurcation Amplifier) has been developed where an underdamped microwave driven JJ is operated in a bistable regime close to a bifurcation of two stable dynamical states \cite{siddiqui_04,metcalfe_07}.
In this range the JJ acts as an anharmonic oscillator which is extremely sensitive to thermal and quantum fluctuations and may thus switch from one dynamical state (small-amplitude oscillation) to the other one (large-amplitude oscillation) or vice versa. Even though in this type of switching process the escape happens to occur across a dynamical rather than an energy barrier, in a moving frame picture rotating with the frequency of the external driving force, the situation can be mapped on a time independent escape problem with a non-standard Hamiltonian \cite{marthaler_06}.

Accordingly, a classical rate theory has been developed already two decades ago \cite{dykman_90}. It is founded on
 Kramers' seminal approach \cite{kramers_40} to calculate the decay rate from the steady solution of a corresponding time evolution equation for the probability density. This approach has been substantially extended in the 1980s \cite{haenggi_90},
 in particular, to lower temperatures where below a so-called crossover temperature quantum tunneling prevails
against thermal activation  \cite{caldeira_81,grabert_84}.  In between there is a broad range where both quantum and thermal fluctuations are
strongly intermingled. The details of this behavior depend very much on the friction strength though \cite{leggett_92,weiss_07}.
 Qualitatively, two regimes must be distinguished \cite{haenggi_90}, namely, the range of moderate to strong friction, where classically the bottleneck for escape is spatial diffusion, and the range of weak friction, where the escape is controlled by energy diffusion. In the former one, typically realized in condensed phase systems, quantum corrections have been derived within dynamical formulations as well as within thermodynamical approaches \cite{grabert_87,ankerhold_07}.

More involved is the situation in the energy diffusive regime, found in gas phase systems like e.g.\ ensembles of atoms or molecules, since at lower temperatures an appropriate time evolution equation in energy space is not at hand a priori.
In some approaches the classical energy diffusion equation has been extended {\em ad hoc} by adding a loss term which captures the finite transmission through the potential barrier near its top \cite{rips_86,dekker_88}. The continuum version of a quantum mechanical master equation has been the starting point for other studies \cite{griff_89,rips_90,coffey_07}, where transition matrix elements have been treated classically, however, while finite transmission {\em and} reflection probabilities appear as quantum mechanical input. It was shown that quantum fluctuations lead to an intricate behavior for low friction and that in certain ranges of parameter space the reflection near the barrier top may even prevail so that the escape rate shrinks below its classical value. Classical transition probabilities have also been used in \cite{larkin_86} and inserted in a discrete version of a master equation. In
\cite{marthaler_06} a master equation strictly valid only for harmonic systems has been applied to study quantum effects in the switching between dynamical states as described above.
Hence, what is really missing, is a consistent semiclassical derivation of an energy diffusion equation for escape out of a metastable well in the regime of intermediate temperatures (above crossover) from a general master equation. Note that below the crossover the energy diffusive domain ceases to exist quickly with decreasing temperature.  The main difference to previous treatments is that quantum effects like tunneling and reflection are treated systematically and thus must also be taken into account in transition matrix elements, at least for energies near the barrier top.  This way, we obtain a semiclassical generalization of the classical energy diffusion equation which will be used in this paper to analyze the impact of quantum fluctuations on the escape for a standard model of an energy barrier. In a subsequent publication the situation described above, namely, quantum effects in the switching between dynamical states will be treated.

The paper is organized as follows: In Sec.~\ref{classical} we recall the classical results and introduce the basic notations. In Sec.~\ref{sc:qm} it is shown how to re-derive the classical energy diffusion equation from a semiclassical limit to the quantum master equation, where tunneling is neglected. The latter process is taken into account systematically in Sec.~\ref{finiteT}, where the main result, namely, a quantum version of the classical
energy diffusion equation is derived. Quantum corrections to the escape rate for a standard potential model are obtained in Sec.~\ref{qrate} and discussed in Sec.~\ref{discuss}.

\section{Classical regime}\label{classical}
 In the standard description of dissipative systems the total Hamiltonian is of the form \cite{caldeira_83,weiss_07}
\begin{eqnarray}
 H&=&H_S+H_B+H_{SB}\nonumber\\
&=&\frac{p^2}{2m}+V(q)+\sum_{n=1}^N\frac{p_n^2}{2m_n}\nonumber\\
&+&\frac{m_n}{2}\omega_n^2 x_n^2-q\sum_{n=1}^N c_nx_n+q^2\sum_{n=1}^N\frac{c^2_n}{2 m_n \omega_n^2}\, ,
\end{eqnarray}
where $H_S$ contains the system part with a metastable potential
\begin{equation}
\label{pot}
 V(q)=\frac{1}{2}m\omega_0^2q^2\left(1-\frac{q}{q_0}\right),
\end{equation}
with a well region around $q=0$ separated from a continuum by an energy barrier of
height $\Delta V=\frac{2}{27}m \omega_0^2 q_0^2$. The heat bath $H_B$ consists of a large number of harmonic oscillators linearly coupled via $H_{SB}$ to the system.

To setup the stage and for later purposes we first consider the well-known classical realm.
 There, after integrating out the bath degrees of freedom the associated equation of motion for the system coincides with a generalized Langevin equation \cite{zwanzig_73}
\begin{equation}
 m \ddot{q}(t)+\frac{\partial V}{\partial q}+m\int_0^t ds \hat{\gamma}(t-s) \dot{q}(s)=\xi(t).
\end{equation}
Here  the influence of the heat bath at temperature $T=1/k_{\rm B}\beta$ is described by a damping kernel $\hat{\gamma}(t)$ which is related to the Gaussian noise force $\xi(t)$ with vanishing mean via the dissipation fluctuation relation
\begin{equation}
 \langle\xi(s)\xi(t)\rangle=\frac{m}{\beta}\hat{\gamma}(|s-t|)\, .
\end{equation}
In the sequel we will assume that the bath memory time is much shorter than other relevant time scales so that a Markovian kernel of the form $\hat{\gamma}(t)=2 \gamma \delta(t)$ applies.

Thermally activated decay in the regime of weak dissipation requires that the energy lost by a classical particle during one period of oscillation in the metastable well is sufficiently smaller than the thermal energy. Specifically, as first shown by Kramers \cite{kramers_40} the condition reads
\begin{equation}
\gamma I(E=\Delta V)\ll1/\beta
\end{equation}
with the action $I(E)$ along a periodic orbit being almost constant for energies close to the barrier top, which is the relevant energy range for the escape process.
A convenient procedure to derive the rate for this process converts the above Langevin dynamics in an equivalent equation of motion for the probability distribution $W(E,\phi,t)$ to find orbits with energy $E$ and phase $\phi$ at time $t$ in the well. Since for low friction energy is almost conserved, while the corresponding phase oscillates fast, the latter one can be adiabatically eliminated and one arrives at
 the energy-diffusion equation for the marginal probability $P(E,t)$ \cite{kramers_40}
\begin{equation}\label{eq:e_dif_c}
 \dot{P}(E,t)=\frac{\partial}{\partial E}\Delta(E)\left(1+\frac{1}{\beta}\frac{\partial}{\partial E}\right)\frac{\omega(E)}{2\pi}P(E,t)\, .
\end{equation}
Here,  $\Delta(E)$ is the energy relaxation coefficient
\begin{equation}
\Delta(E)=\gamma I(E)\, .
\end{equation}
The escape rate is determined by the stationary nonequilibrium distribution $P_{\rm st}(E)$ to (\ref{eq:e_dif_c}) which is associated with a finite flux across the barrier and obeys the boundary conditions that $P_{\rm st}=0$ for $E>\Delta V$ and that $P_{\rm st}(E)$ approaches a Boltzmann distribution in the well region. Accordingly, one obtains for high barriers $\beta \Delta V\gg 1$ the classical Kramers result
\begin{eqnarray}
\label{classrate}
\Gamma_{\rm cl}&=&\int_0^\infty dE \, P_{\rm st}(E) \nonumber\\
&=&\frac{\omega_0 \gamma \, I(\Delta V)\beta}{2\pi}\, {\rm e}^{-\beta \Delta V}.
\end{eqnarray}

The above expression can also be derived by working in a multidimensional space including the system degree of freedom and the bath oscillators. One introduces normal mode coordinates in the parabolic range around the barrier top, where the total system is separable, and then studies the dynamics of the {\em unstable} normal mode in presence of the coupling to the stable ones due to the potential anharmonicity. This methodology allows to capture the full turnover from the regime of weak dissipation considered here to the regime of spatial diffusion for higher friction \cite{pollak_86_A,pollak_89_A}. For weak dissipation one recovers to leading order (\ref{classrate}), while an improved result  including higher order corrections (but still away from the turnover region) reads $\Gamma_{\rm cl, imp}=(\lambda_{\rm b}/\omega_{\rm b})\, \Gamma_{\rm cl}$. Here, finite recrossing at the barrier top is described by the Grote-Hynes frequency \cite{grote_80}
\begin{equation}
\label{grote}
\lambda_{\rm b}=\sqrt{\frac{\gamma^2}{4}+\omega_{\rm b}^2}-\frac{\gamma}{2}\, ,
\end{equation}
which describes the effective barrier frequency of the unstable normal mode, while $\omega_{\rm b}$ is the bare barrier frequency. For the model (\ref{pot}) one has $\omega_b=\omega_0$.

\section{Semiclassical derivation of the Kramers equation}\label{sc:qm}
In a first step and as test-bed for the semiclassical approximation we derive here the classical Kramers equation (\ref{eq:e_dif_c}) from the Pauli-master equation. For this we follow the approach developed in \cite{karrlein_97} for potentials with stable bound states.
Thus, we assume a number of $N$ discrete energy levels in the well $E\leq \Delta V$ and at this point neglect quantum tunneling through the barrier completely (cf.~fig.~\ref{fig:rate1}). At the end, $N$ will be taken to be very large so that one can work with a quasi-continuum of states.
Accordingly, we have
\begin{eqnarray}
 H_S|n\rangle=E_n|n\rangle,& & n=1,2...N
\end{eqnarray}
with the normalization
\begin{equation}\label{eq:norma1}
 \langle n|m\rangle=\delta_{n,m}.
\end{equation}
Now, let $p_n(t)$ be the probability to find the system in discrete state $n$ at time $t$. Then, one can derive from the exact path integral expression of the reduced density matrix for weak coupling to the heat bath and not too low temperatures Pauli's master equation, i.e.,
\begin{equation}\label{eq:master_d}
\dot{p}_n(t)=\sum_{m=0}^N[W_{n,m}p_m(t)-W_{m,n}p_n(t)].
\end{equation}
The transition rates from state $m$ to $n$ are given by \cite{breuer_02}
\begin{equation}
W_{n,m}=\frac{1}{\hbar^2}\int_{-\infty}^{\infty}dt \mbox{Tr}_B[\langle n|H_{SB}(t)|m\rangle\langle m|H_{SB}|n\rangle\rho_B^{eq}]
\end{equation}
 with $\mbox{Tr}_B$ denoting the trace over the bath, the coupling in the interaction picture $H_{SB}(t)=e^{i(H_S+H_B)t/\hbar}H_{SB}e^{-i(H_S+H_B)t/\hbar}$, and $\rho_B^{eq}={\rm e}^{\beta H_B}/\mbox{Tr}_B[{\rm e}^{\beta H_B}]$ the equilibrium bath density matrix. The above expression can be evaluated explicitly in case of the linear system-bath coupling and one arrives at the golden rule type of formula
\begin{equation}\label{eq:Wl}
 W_{n,m}=\frac{1}{\hbar^2}|\langle n|q|m\rangle|^2D(E_n-E_m)
\end{equation}
with the bath absorption/emission captured by
\begin{equation}
\label{bathdensity}
 D(E_n-E_k)=2m\gamma (E_n-E_k)\, \overline{n}(E_n-E_k)\, ,
\end{equation}
where
\begin{equation}
 \overline{n}(E_n-E_k)=\frac{1}{e^{\beta(E_n-E_k)}-1}.
\end{equation}

The probability $p_n(t)$ is related to the probability $P(E,t)$ to find the system in a state with energy between $E$ and $E+dE$ at time $t$ via
\begin{equation}\label{eq:pe}
P(E,t)=\sum_{n=0}^N\delta(E-E_n)p_n(t)
\end{equation}
which allows to reformulate the above master equation (\ref{eq:master_d}) as
\begin{eqnarray}\label{eq:master_d2}
\hspace{-.5cm}\dot{P}(E,t)=\sum_{m,n=0}^N\delta(E-E_n)\left[W_{n,m}p_m(t)-W_{m,n}p_n(t)\right]\nonumber\\
\hspace{-.5cm}=\sum_{n=0}^N\sum_{l=-n}^{N-n}\left[\delta(E-E_{n+l})-\delta(E-E_n)\right] W_{n+l,n}p_n(t).
\end{eqnarray}

\subsection{Semiclassical wave function}\label{sc:swf}

To calculate the transition matrix elements in system space we now apply the semiclassical WKB approximation \cite{karrlein_97}.
Up to second order in $\hbar$ one  obtains for the wave functions
\begin{equation}
\label{wave1}
\langle n|q\rangle\equiv\langle E_n|q\rangle=\frac{1}{2}\left[ \langle E_n|q\rangle^-+\langle E_n|q\rangle^+\right]
\end{equation}
with
\begin{equation}\label{eq:wave}
\langle E|q\rangle^{\pm}=\frac{N(E)}{\sqrt{p(E,q)}}\, {\rm e}^{\pm\frac{i}{h}S_0(E,q)\mp \frac{i \pi}{4}}
\end{equation}
and the action $S_0(E,q)=\int_{q_1}^{q}p(E,q^{\prime})dq^{\prime}$ of an orbit starting at $q_1$ and running in time $t$ towards $q$ with momentum $p(E,q^\prime)$ at energy $E$ \cite{dunham_32}.
The normalization is determined from (\ref{eq:norma1})
\begin{equation}
N(E)=\sqrt{2m \omega(E)/\pi}
\end{equation}
with $\omega(E)$ being the frequency of the classical oscillation at energy $E$.
Note that forward and backward wave contributions in (\ref{eq:wave}) are related by
\begin{equation}
\left[\langle E|q\rangle^+\right]^* =\langle E|q\rangle^-.
\end{equation}
The energy $E_n=E$ of the discrete spectrum is given by the quantization condition
\begin{equation}\label{eq:quant}
 \frac{1}{2 \pi \hbar}\oint dq p(E,q)=n+\frac{1}{2}
\end{equation}
containing the action over periodic orbits in the well.

\subsection{Semiclassical matrix elements}\label{sc:smed}
With the semiclassical wave function at hand we can now calculate the transition matrix elements which enter the transition rates in (\ref{eq:Wl}).
According to the restricted interference approximation \cite{more_91} we only keep the diagonal contributions of forward/backward waves to obtain
\begin{eqnarray}\label{eq:int}
\lefteqn{\langle E_m|q|E_n\rangle}\nonumber\\
&&\simeq\frac{1}{4}\int_{q_1}^{q_2} dq\, q \left[\langle E_m|q\rangle^+\langle E_n|q\rangle^-+\langle E_m|q\rangle^-\langle E_n|q\rangle^+\right]\nonumber\\
&&=\frac{1}{4}\oint dq \langle E_m|q\rangle^+q\langle E_n|q\rangle^-\equiv Q_{\rm scl}^{(n,m)}\, ,
\end{eqnarray}
where $q_1, q_2$ denote the left and the right turning points, respectively, of the periodic orbit with energy $E$.
To calculate $Q_{\rm scl}^{(n,m)}$ we exploit that due to $[H_S,q]=\hbar p/(i m)$ the matrix elements $Q^{(n,m)}=\langle E_m|q|E_n\rangle$ and
$P^{(n,m)}=\langle E_m|p|E_n\rangle$ are related via
\begin{equation}\label{eq:pq}
 P^{(n,m)}=\frac{i m}{\hbar}(E_m-E_n)Q^{(n,m)}\, .
\end{equation}
This way one finds together with (\ref{eq:wave})
\begin{eqnarray}
\label{qcl}
Q_{\rm scl}^{(n,m)}=\frac{\hbar}{2 \pi i( E_m-E_n)}\oint dq\sqrt{\frac{\omega(E_n)\omega( E_m)}{p(E_n)p(E_m)}}\nonumber\\
\left[p(E_n)+\frac{\hbar}{2 i}\frac{1}{p(E_n)}\frac{\partial p}{\partial q}\right]e^{\frac{i}{\hbar}[S_0(E_n,q)-S_0(E_m,q)]}
\end{eqnarray}
The exponential is further simplified by applying  the semiclassical expansion of energy differences based on (\ref{eq:quant}), i.e.,
\begin{equation}
 \hbar A_l(E):=E_{n+l}-E_n=\hbar l\omega(E_n)+\frac{\hbar^2}{4}l^2[\omega(E_n)^2]^{\prime}+o(\hbar^3)\, ,
\end{equation}
where here and in the sequel the prime $^\prime$ at energy dependent functions denotes the derivative with respect to energy.
Now, $Q_{\rm scl}^{(l)}(E_n)\equiv Q_{\rm scl}^{(n,m)}, m-n=l$ can be systematically expanded in powers of $\hbar$ based on $S_0(E_n,q)-S_0(E_m,q)\approx -\hbar l \omega(E_n) t(E_n,q)+\frac{\hbar^2}{2}l^2\omega(E_n)(\omega(E_n)t(E_n,q))^{\prime}$ with $t(E,q)$ being the time a segment of a periodic orbit with energy $E$ needs to reach  position $q$ from its turning point.
The expansion for the squared matrix element is thus found from (\ref{qcl}) to read \cite{karrlein_97}
\begin{equation}\label{eq:qd}
|Q_{\rm scl}^{(l)}|^2={Q_{\rm cl}^{(l)}}^2+\frac{\omega\hbar l}{2}\left[{Q_{\rm cl}^{(l)}}^2\right]^{\prime}\, ,
\end{equation}
with
\begin{equation}\label{eq:q}
Q_{\rm cl}^{(l)}(E)=\frac{1}{2i \pi l}\oint dq e^{-i l\omega(E)t(E,q)}
\end{equation}
describing the classical transition amplitude.
From this finding the semiclassical expansion of the transition rates (\ref{eq:Wl}) between states with energy $E_n$ and energy $E_m$
such that $E_n-E_m=l \hbar \omega(E_n)$ is given by
\begin{eqnarray}\label{eq:w_d}
 \lefteqn{W_{n,m}=W_l(E_n)=}\nonumber\\
&&\frac{2m\gamma}{\beta\hbar^2}\, {Q_{\rm cl}^{(l)}}^2+\frac{lm\omega\gamma}{\hbar\beta}\,\left(-\beta {Q_{\rm cl}^{(l)}}^2+[{Q_{\rm cl}^{(l)}}^2]^{\prime}
\right)\, .
\end{eqnarray}

\subsection{Semiclassical expansion of the master equation}\label{sc:seme}

The above results are now used to determine the semiclassical expansion of the master equation
(\ref{eq:master_d2}), which in turn means to treat a large number of quasi-stationary energy eigenstates in the well as a quasi-continuum. After some straightforward manipulations (see Appendix~\ref{appA}) one arrives at
\begin{eqnarray}
\label{hbardiffus}
\dot{P}(E,t)&=&\sum_{k=1}^{\infty}\left(\frac{\partial}{\partial E}\right)^k\frac{(-\hbar)^k}{k!}\nonumber\\
&\times&\sum_{l=1}^N\left[W_l(E)A_l^k(E)+W_{-l}(E)A_{-l}^k(E)\right]P(E,t)\nonumber\\
&&
\end{eqnarray}
In case of a harmonic system in the second sum only terms with $l= 1$ contribute and together with the energy independence of its frequency $\omega_0$ the first sum gives rise to differential operators of the form $\exp(\pm \hbar\omega_0 \partial/\partial E)$. For anharmonic systems we proceed in the spirit of the semiclassical expansion and take into account only terms up to $k=2$, i.e.,
\begin{eqnarray}
\dot{P}(E,t)=\frac{\partial}{\partial E}\left[-D_1 + \frac{\partial}{\partial E}D_2\right]P(E,t)
\end{eqnarray}
with
\begin{equation}
D_k(E)=\sum_{l=-N}^N \frac{\hbar^k}{k!}\ W_l(E)A_l^k(E)\, .\label{eq:B}
\end{equation}
Combining this result with (\ref{eq:w_d}) provides the energy diffusion equation in leading order in $\hbar$, namely,
\begin{equation}
\dot{P}(E,t)=\frac{\partial}{\partial E}\Delta\left[1+\frac{1}{\beta} \frac{\partial}{\partial E }\right]\frac{\omega}{2\pi}P(E,t)\, ,
\label{diffusion1}
\end{equation}
where
\begin{equation}\label{eq:sum}
 \Delta(E)=2m\omega(E)\gamma \pi\sum_{l=-N}^N l^2\, Q_{\rm cl}^{(l)}(E)^{2}
\end{equation}
In the limit of a quasi-continuum of states ($N\gg 1$) this sum can be recast with the help (\ref{eq:q})
as
\begin{equation}
 \Delta(E)=\gamma\oint dq \, p(E,q)=\gamma I(E)
\end{equation}
with the action of a periodic orbit with energy $E$
\begin{equation}
\label{periodicaction}
I(E)=\oint dq p(E,q)\, .
\end{equation}
Accordingly, we recover from (\ref{diffusion1}) the classical Kramers equation (\ref{eq:e_dif_c}), i.e.\
 $\dot{P}(E,t)={\mathcal L}_E^{(\rm cl)}\, P(E)$ with
\begin{equation}\label{eq:Ld}
 \mathcal L_E^{(\rm cl)}=\frac{\partial}{\partial E}\gamma I(E)\left(1+\frac{1}{\beta}\frac{\partial}{\partial E}\right)\frac{\omega(E)}{2\pi}\, .
\end{equation}
The main result of this derivation is that up to second order in the $\hbar$-expansion (\ref{hbardiffus}) quantum effects in the energy diffusion do not appear. Of course, the above procedure can be elaborated to systematically include higher order $\hbar$ correction. We will see in the sequel though, that for the barrier crossing problem, finite reflection and transmission amplitudes provide much larger contributions, which are formally of order 1.

\section{Energy diffusion for finite transmission}\label{finiteT}

\begin{figure}[htbp]
\vspace*{0.1cm}
\epsfig{file=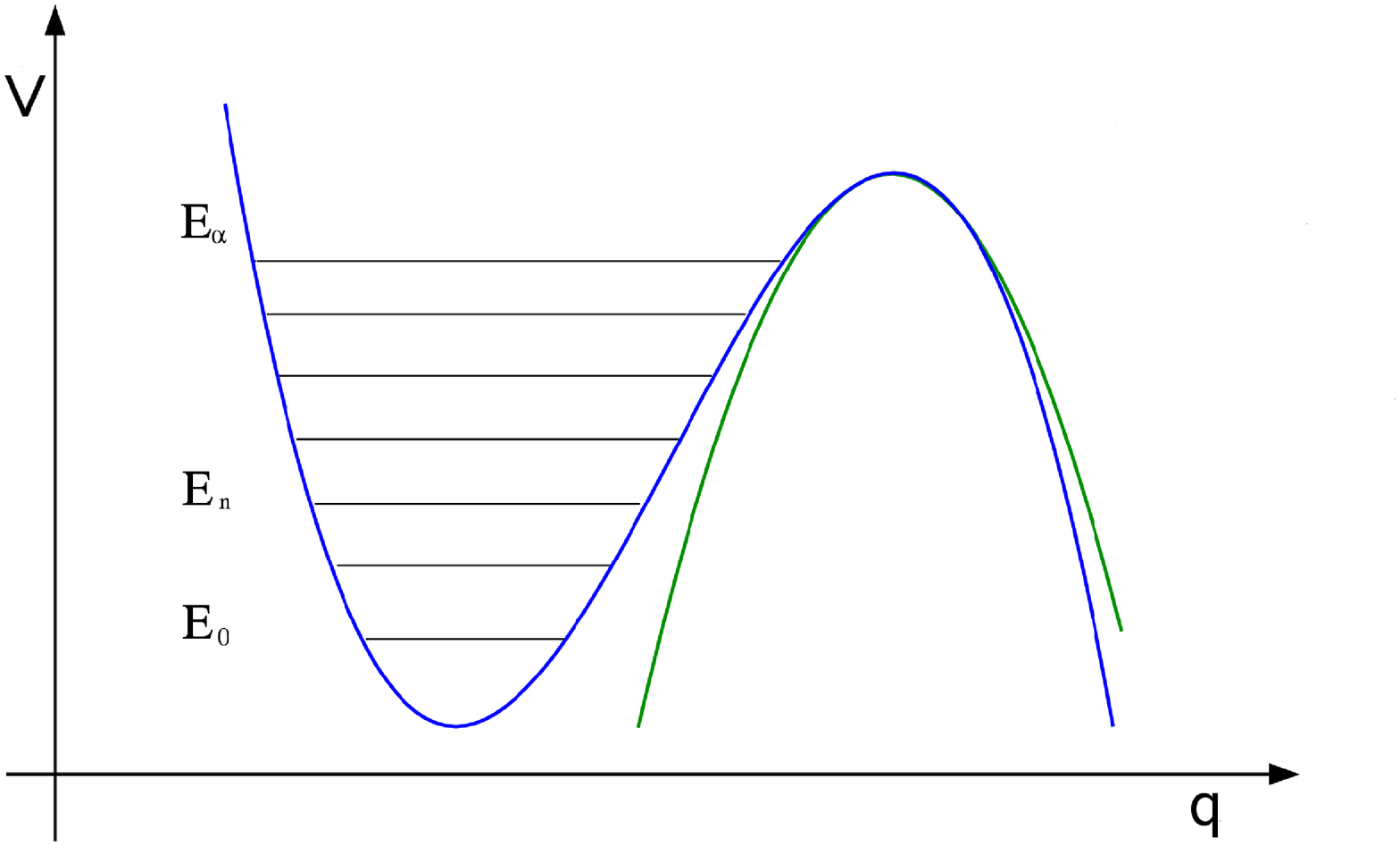, height=5.3cm}
\caption{Typical metastable well potential discussed in the text with quasi-stationary states deep in the well and a continuum of states near the barrier top, around which a parabolic barrier approximation applies.}
\label{fig:rate1}
\vspace*{-0.25cm}
\end{figure}

In the previous section the semiclassical wave functions (\ref{wave1}), (\ref{eq:wave}) were taken as stationary eigenstates in the well potential for all energies below the barrier top. The underlying reasoning that finite barrier transmission does not play a role, however, fails as the temperature is lowered. Namely, for energies $E<\Delta V$ the finite probability $T(E)$ to tunnel through the barrier and  for energies $E>\Delta V$ the finite probability $R(E)$ to be reflected from the barrier can only be neglected as long as near the barrier top the thermal energy scale  $k_B T$ by far exceeds the quantum scale $\hbar\omega_{\rm b}$. When this does no longer apply the consequences are two-fold: On the one hand in a range somewhat below $\Delta V$ discrete energy levels are strongly broadened due to tunneling so that one has a continuum of states in the well strongly coupled to the continuum on the right side of the barrier (see fig.\ref{fig:rate1}); on the other hand, continuum eigenstates with $E>\Delta V$ gain a longer life-time in the well region due a finite reflection and may thus influence the escape process as well. Note that corresponding quantum effects are substantial since $R(\Delta V)=T(\Delta V)=1/2$, and thus dominate against higher order $\hbar$-corrections (associated with higher than second order derivatives in $E$) in the expansion (\ref{hbardiffus}) \cite{karrlein_97}. However, it is well-known that in the limit $\hbar\to 0$ the functions $R$ and $T$ are non-analytical in $\hbar$ [see (\ref{eq:dif1})] so that an expansion in powers of $\hbar$ is not feasible.
In a semiclassical approach we thus proceed to work  in a continuum representation right from the beginning and  treat low lying states and higher lying states on equal footing. This approach is justified as long as temperature is not too low and tunneling prevails near the top of the barrier.
Hence, instead of (\ref{eq:pe}) we now start with
\begin{equation}
 P(E,t)=\int_{-\infty}^{\infty} p_{E^{\prime}}(t)\delta(E-E^{\prime})dE^{\prime}.
\end{equation}
Then, $R(E)\, p_E(t) dE$ is the probability to find the system at time $t$ in a state with an energy between $E$ and $E+dE$ inside the well and $T(E)\, p_E(t)$ with $T(E)=1-R(E)$ the probability for tunneling out of the well.
In case of finite barrier transmission apart from transitions for states inside well induced by the heat bath, there is also a loss of population due to tunneling through the barrier.
Hence, the variation in the probability  $p_E(t)$ to find the system in a state with energy $E$ in the well is equal to the probability that the system arrives in the well from an other energy $E^{\prime}$ minus the probability that the system leaves the state with energy $E$ to another state in the well region and minus the probability of tunneling out of the well, i.e.,
\begin{eqnarray}
\dot{p}_E(t)&=&\int_{}dE^{\prime}\left[W_{E,E^{\prime}}\frac{R(E^{\prime})p_{E^{\prime}}(t)}{n(E^{\prime})}\right.\\
&&\left. -W_{E^{\prime},E}\frac{R(E)p_E(t)}{n(E)}\right]-T(E)\frac{\omega(E)}{2 \pi}p_E(t),\nonumber
\end{eqnarray}
where $n(E)$ is the density of states. Now, in generalization of (\ref{eq:master_d2}) we have
\begin{eqnarray}
\label{PE}
\dot{P}(E,t)&=&\int_{}dE^{\prime}dE^{\prime\prime}\left[W_{E^{\prime\prime},E^{\prime}}\frac{R(E^{\prime})
p_{E^{\prime}}(t)}
{n(E^{\prime})}\right.\nonumber\\
&&\left.-W_{E^{\prime},E^{\prime\prime}}\frac{R(E^{\prime\prime})p_{E^{\prime\prime}}(t)}
{n(E^{\prime\prime})}\right]\delta(E-E^{\prime\prime})\nonumber\\
&&-T(E)\frac{\omega(E)}{2 \pi}p_{E}(t)
\end{eqnarray}
with $W_{E,E^{\prime}}$ being the transition rate from state with energy $E$ to state with energy $E^{\prime}$, namely,
\begin{eqnarray}\label{eq:Wq}
W_{E,E^{\prime}}&=&\frac{1}{\hbar^2}\int_{-\infty}^{\infty}dt Tr_B[\langle E|H_{SB}(t)|E^{\prime}\rangle\langle E^{\prime}|H_{SB}|E\rangle\rho_B^{eq}]\nonumber\\
&=&\frac{1}{\hbar^2}|Q_{qm}(E',E)|^2D(E-E^{\prime}),
\end{eqnarray}
where $Q_{\rm qm}(E',E)\equiv\langle E^{\prime}|q|E\rangle$.
Here, the bath spectral function $D(E)$ is defined according to (\ref{bathdensity}). The semiclassical wave functions entering the transition matrix elements for the system degree of freedom, however, must now include the finite transmission/reflection probability.
In the energy range close to the barrier top, where tunneling dominates in the temperature range considered,  the WKB approximation is not applicable because the classical turning points to the left and to the right of the barrier are not sufficiently separated. In this situation, one exploits that the Schr\"odinger equation for a parabolic barrier with barrier frequency $\omega_{\rm b}$ can be solved exactly. The proper eigenfunctions are then matched asymptotically (sufficiently away from the barrier top) onto WKB wave functions to determine phases and amplitudes of the latter ones.
This way, one obtains
\begin{eqnarray}\label{eq:m}
\langle E|q\rangle=\frac{1}{2}\tilde{N}(E)\left[ \langle E|q\rangle^-+r(E) \langle E|q\rangle^+\right]
\end{eqnarray}
with the matrix elements (\ref{eq:wave}) and  the complex valued reflection amplitude $r(E)$ of a parabolic barrier \cite{atabek_91} related to the reflection probability $R(E)=|r(E)|^2$. The normalization follows from $\langle E|E'\rangle=\delta(E-E')$ as
\begin{equation}
 \tilde{N}(E)=\frac{1}{\sqrt{\hbar\omega}}\sqrt{\frac{2}{1+R(E)}}\, .
\end{equation}
In case of vanishing transmission, $E<<\Delta V$ and  $R\to 1$, one recovers the previous result $\tilde{N}(E)\to 1/\sqrt{\hbar \omega}$ so that it is possible to use (\ref{eq:m}) for  all energies provided the length scale where a parabolic approximation for the barrier applies is much larger than the quantum mechanical length scale $\sqrt{\hbar/m\omega_{\rm b}}$. This in turn requires a high barrier and thus follows directly from the condition for metastability.

\subsection{Semiclassical matrix elements}
In a similar way as in Sec.~\ref{sc:smed} we now calculate in the semiclassical limit the transition matrix elements $ Q_{\rm qm}(E',E)$  for finite transmission through the barrier. One has
\begin{eqnarray}
Q_{\rm qm}(E',E)&\simeq&\frac{\tilde{N}(E^{\prime})\tilde{N}(E)}{4}\int_{q_1}^{q_2}dq \left[\langle E^{\prime}|q\rangle^+q\langle E|q\rangle^-\right.\nonumber\\
&&\hspace{1cm}\left.+ r(E)r(E^{\prime})^*\langle E^{\prime}|q\rangle^-q\langle E|q\rangle^+\right]\nonumber\\
&=&\tilde{N}(E^{\prime})\tilde{N}(E)\left[Q_{\rm scl}^*+r(E)r(E^{\prime})^* Q_{\rm scl}\right]\, , \nonumber\\
\end{eqnarray}
where in generalization of (\ref{eq:int})
\begin{eqnarray}
Q_{\rm scl}(E',E)=\frac{1}{4}\int_{q_1}^{q_2}
 dq \langle E'|q\rangle^- q\langle E|q\rangle^+\, .
\end{eqnarray}
Note that in contrast to $Q_{\rm scl}^{(n,m)}$, which is real, the transition element $Q_{\rm qm}$ is complex for finite tunneling amplitudes $|r|>0$. Hence, $Q_{\rm qm}$ cannot be written as a loop integral as in (\ref{eq:int}).
To proceed, one introduces
\begin{equation}
 E^{\prime}-E\equiv\delta\, .
\end{equation}
Now, exploiting again (\ref{eq:pq}) one has
\begin{eqnarray}\label{eq:expQ}
Q_{\rm scl}^{(\delta)}=Q_{\rm cl}^{(\delta)}+\delta\, \left(\frac{1}{2} \left[Q_{\rm cl}^{(\delta)}\right]^{'}+K^{(\delta)}\right)\, ,
\end{eqnarray}
where
\begin{eqnarray}\label{eq:qc}
Q_{\rm cl}^{(\delta)}(E)=\frac{\hbar \omega(E)}{2 \pi i\, \delta}\int_{q_1}^{q_2} dq\  {\rm e}^{-i t(E,q) \delta/\hbar}
\end{eqnarray}
and
\begin{eqnarray}\label{eq:k}
K^{(\delta)}(E)&=&\frac{\hbar\omega(E)}{4\pi i \, \delta}\left\{\int_{q_1}^{q_2} dq {\rm e}^{-i t(E,q) \delta/\hbar}\left[- \frac{p^{\prime}(q)}{p(q)}\right.\right.\\
&+&\left.\left.\frac{\hbar}{i p(q)^2\, \delta}\frac{\partial p(q)}{\partial q}\right]
+2\left[q_2'-q_1'{\rm e}^{-i\pi\delta/\hbar\omega(E)}\right]\right\}\,.\nonumber
\end{eqnarray}
This latter contribution is a boundary term which vanishes for the closed integral in (\ref{eq:q}).
The $\hbar$-expansion of the squared matrix element thus reads
 \begin{equation}\label{eq:w_c}
|Q_{\rm qm}|^2= \tilde{N}^4 \tilde{A}_{\rm qm} + \delta \left( \tilde{B}_{\rm qm} + \tilde{N}^4 \tilde{C}_{\rm qm}\right)
\end{equation}
with coefficients $\tilde{A}_{\rm qm}$, $\tilde{B}_{\rm qm}$, and $\tilde{C}_{\rm qm}$ specified in Appendix~\ref{appB}.

\subsection{Expansion of the master equation}

We now proceed along the same way as done for a discrete spectrum in Sec.~(\ref{sc:seme}) with the substitutions $\hbar A_l\rightarrow\delta$ and $\sum_l\delta\rightarrow\int d\delta $. Thus, we obtain from (\ref{PE}) the expansion
\begin{eqnarray}
\label{expansionPE}
\lefteqn{\dot{P}(E,t)=\sum_{k=1}^{\infty}\left(\frac{\partial}{\partial E}\right)^k\frac{1}{k!}\int_{-\infty}^{\infty}d\delta\, W_\delta(E)\ (-\delta)^k}\nonumber\\
&&\hspace{1cm}\times  \frac{R(E)\, P(E,t)}{n(E)}-T(E)\frac{\omega(E)}{2 \pi}P(E,t)
\end{eqnarray}
with $W_\delta(E)=W_{E,E'}$ for $E'-E=\delta$.
The leading order terms in the sum above with $k=1$ and $k=2$  are kept to get the energy diffusion equation for finite transmission in the semiclassical limit, i.e.\  $\dot{P}(E,T)={\mathcal L}_E^{(\rm scl)} P(E,t)$, where
\begin{equation}\label{eq:L_scl}
{\mathcal L}_E^{(\rm scl)}=\frac{\partial}{\partial E}\left[-\langle\delta \rangle+\frac{\partial}{\partial E} \langle\delta^2\rangle\right]R(E)-T(E)\frac{\omega(E)}{2 \pi}\, .
\end{equation}
Here, the moments of the energy fluctuations read
\begin{equation}\label{eq:epsilon}
 \langle\delta^k\rangle=\frac{1}{n(E)}\int_{-\infty}^{\infty}d\delta W_{\delta}(E)\frac{\delta^k}{k!}
\end{equation}
Apparently, when comparing this expression with (\ref{eq:B}) one observes that drift and diffusion coefficients $D_k$ correspond, as expected, to $\langle \delta^k\rangle=\langle (E'-E)^k\rangle$ for $k=1, 2$, respectively. Now, combining the expansion for the density of states
\begin{equation}\label{eq:density}
n(E)=\frac{1}{\hbar\omega(E)}+O(\hbar)
\end{equation}
with Eqs.~(\ref{eq:expQ})-(\ref{eq:w_c}) one finds upon evaluating the moments (\ref{eq:epsilon}) the first main result, namely, the  semiclassical expression (for details see Appendix~\ref{appC}) of the evolution operator in the energy diffusive regime $\dot{P}(E,t)=\mathcal L_E^{(\rm scl)}P(E,t)$, i.e.,
\begin{eqnarray}
\label{LEQM}
\mathcal L_E^{(\rm scl)} &=&\frac{\partial}{\partial E}\ C(E)\, \gamma I(E)\left(1+\frac{1}{\beta}\frac{\partial}{\partial E}\right)\frac{\omega(E) R(E)}{2 \pi }\nonumber\\
&-&T(E)\frac{\omega(E)}{2 \pi}\, ,
\end{eqnarray}
with
\begin{equation}
C(E)=2 \frac{1+R(E)^2}{[1+R(E)]^2}\, .
\end{equation}
Some remarks are in order here: First, for vanishing transmission ($R=1, T=0$) one recovers from the above expression the classical diffusion operator (\ref{eq:Ld}). This in turn proves what we have said above, namely, that on this level of semiclassical expansion an energy diffusion operator can be derived starting either from a discrete or a continuous spectrum in the well. The difference is though, that the latter procedure conveniently accounts for barrier tunneling near the barrier top with the property that ${\mathcal L}_E^{(\rm scl)}\to {\mathcal L}_E^{(\rm cl)}$ in the range $E\ll \Delta V$, where $R(E)\to 1$.
Second, higher order $\hbar$ corrections in the expansion (\ref{expansionPE}) can now be calculated accordingly, where, however, in case of a barrier crossing problem corresponding contributions are much smaller than those originating from quantum transmission and reflection.
Third, in various previous works  an operator similar to the above one but with the factor $C(E)=1$ has been used to study quantum effects in thermal activation in the low damping regime \cite{melnikov_85,rips_90,dekker_88,griff_89}. The important difference is though that here we derived this operator including its transition matrix elements within a consistent semiclassical expansion of the underlying wave functions, while in the former cases in (\ref{PE}) the corresponding {\em classical} expressions were adopted. Hence, the above equation is the systematic generalization of the classical Kramers' equation in energy space to lower temperatures.

\section{Quantum escape rate} \label{qrate}
The escape rate in the quantum regime can now be evaluated similar as in the classical case by searching for the quasi-stationary energy distribution $P_{\rm st}(E)$. For this purpose, we start by specifying the known  transmission and reflection probabilities in a uniform semiclassical approximation, i.e.,
\begin{eqnarray}\label{eq:dif1}
T(E)=|t(E)|^2=\frac{1}{1+\exp[-S_{\rm e}(E)/\hbar]}\nonumber\\
R(E)=|r(E)|^2=\frac{1}{1+\exp[S_{\rm e}(E)/\hbar]}\, ,
\end{eqnarray}
where $S_{\rm e}(E)$ denotes the Euclidian action of a periodic orbit with energy $E$ oscillating in the inverted barrier potential $-V(q)$. In the energy range near the barrier top, where tunneling dominates in the temperatures range considered here, the action reduces to its vale for a parabolic barrier, namely, $S_{\rm e, pb}(E)=2\pi(E-\Delta V)/\omega_{\rm b}$. The above expressions thus smoothly connect the energy range near top with the low energy range, where $T(E)$ drops exponentially. Hence, deep inside the well the only quantum effects are zero-point fluctuations. The reflection dependent factor $C(E)$ is always larger than 1 which means that leaking out of the wave function by tunneling appears effectively as an increase in energy loss $C(E) \gamma I(E)>\gamma I(E)$ during one cycle of a classical orbit in the well.

Now, in order to explicitly find the steady-state distribution $P_{\rm st}(E)$ it is convenient to work with
a dimensionless energy measured from the barrier top, $\epsilon=(E-\Delta V)\beta$, and to introduce a dimensionless inverse temperature
\begin{equation}
\theta=\hbar\beta\omega_{\rm b}/2\pi\, . \label{temp}
\end{equation}
 Then, one writes  $f(\epsilon)= [\omega(\epsilon)/2\pi]\, R(\epsilon)P_{\rm st}(\epsilon)$ so that from $\dot{P}_{\rm st}=0$ one arrives with (\ref{LEQM}) at
\begin{equation}\label{eq:dif2}
\beta\frac{\partial}{\partial \epsilon}C(\epsilon)\gamma I(\epsilon)\left(1+\frac{\partial}{\partial \epsilon}\right)f(\epsilon)=\frac{T(\epsilon)}{R(\epsilon)}f(\epsilon)\, .
\end{equation}
The boundary conditions here are such that $f(\epsilon)\to 0$ for $\epsilon\gg 1$ and that inside the well ($\epsilon$ some $\theta$ below $\Delta V$) $f(\epsilon)\to f_\beta(\epsilon)$ with the equilibrium distribution
\begin{equation}
 f_\beta(\epsilon)=\frac{1}{2 \pi \hbar Z_0}{\rm e}^{-\epsilon-\beta\Delta V}\, .
\end{equation}
Here, the  quantum partition function in the harmonic well reads
\begin{equation}
Z_0=\frac{1}{\omega_0\hbar\beta}\prod_{n=1}^{\infty}\frac{\nu_n^2}{\nu_n^2+\omega_0^2+\nu_n \gamma}\, ,
\label{partgamma}
\end{equation}
with Matsubara frequencies $\nu_n=2\pi n/\hbar\beta$. For vanishing friction this reduces to the known result
$Z_{00}=1/[2{\rm sinh}(\omega_0\hbar\beta/2)]$.
With
\begin{equation}
 f(\epsilon)=f_\beta(\epsilon)\ g(\epsilon) \label{fg}
\end{equation}
(\ref{eq:dif2}) becomes
\begin{equation}
\label{diffg}
C(\epsilon) g(\epsilon)''+g(\epsilon)'\left[{C(\epsilon)'}-C(\epsilon)\right]=\frac{1}{\rho}\frac{T(\epsilon)}{ R(\epsilon)}\, g(\epsilon)\, ,
\end{equation}
where we also exploited that the action varies smoothly in the relevant energy range around $\epsilon=0$ (i.e.\ in an interval of some $k_{\rm B}T, \hbar\omega_{\rm b}$ below the top) so that with
$I(\epsilon)\simeq I(\epsilon=0)$ one defines $\rho=\beta \gamma I(0)$.

Analytical progress can now only be made by approximating the coefficient $C(\epsilon)$.
For this purpose we assume self-consistently that $g(\epsilon)$ becomes exponentially small in the semiclassical sense for energies $\epsilon$ of order $\theta$ above the barrier top. Then, (\ref{diffg}) needs only to be considered in a relevant range $\epsilon<\theta\cdot\kappa$ with some constant $\kappa>1$ of order 1.
A very accurate approximation in this energy range  is given by
\begin{equation}
C(\epsilon<\theta\cdot\kappa)\approx
1+ \frac{1}{9}\ {\rm e}^{\epsilon/\theta}
\end{equation}
for $\kappa<1.5$. Since the final rate does not depend on the precise value of $\kappa$, we choose $\kappa={\rm ln}(4)$ which ensures that $C(\epsilon)'$ is very well approximated for all energies up to its maximum located at $\epsilon={\rm ln}(4)\, \theta$.
The solution of (\ref{diffg}) is then found in terms of hypergeometric functions as
\begin{eqnarray}
g(\epsilon)&=&_2F_1\left[\frac{1}{2}-\frac{\theta}{2}-a,\frac{1}{2}-\frac{\theta}{2}+a,1-\theta,-
\frac{{\rm e}^{\epsilon/\theta}}{9}\right]\nonumber\\
&+&B{\rm e}^{\epsilon}\   _2F_1\left[\frac{1}{2}+\frac{\theta}{2}-a,\frac{1}{2}+\frac{\theta}{2}+a,1+\theta,-\frac{{\rm e}^{\epsilon/\theta}}{9}\right]\, ,\nonumber\\
\label{solu1}
\end{eqnarray}
with the coefficients
\begin{equation}
B=-\frac{1}{4^{\theta}}\frac{_2F_1\left[\frac{1}{2}-\frac{\theta}{2}-a,\frac{1}{2}-
\frac{\theta}{2}+a,1-\theta,-\frac{4}{9}\right]}{_2F_1\left[\frac{1}{2}+\frac{\theta}{2}-a,
\frac{1}{2}+\frac{\theta}{2}+a,1+\theta,-\frac{4}{9}\right]}
\end{equation}
and
\begin{equation}
a=\sqrt{\frac{\rho(1-\theta)^2+36\theta^2}{4\rho}}
\end{equation}
A straightforward analysis now verifies our assumption that $g(\epsilon)\to 0$ for $\epsilon$ of order $\theta$ above the barrier top.

The semiclassical approximation to the quantum escape rate
\begin{equation}
\Gamma_{\rm qm}=\int_0^{\infty}dE\ T(E)\, \frac{\omega(E)}{2\pi}\, P_{\rm st}(E)\\
\end{equation}
is obtained by exploiting ${\mathcal L}_E^{(\rm scl)} P_{\rm st}=0$ with (\ref{LEQM}) and (\ref{fg}) as
\begin{equation}
\Gamma_{\rm scl}=-\left. \frac{1}{\beta}\,f_\beta(\epsilon)\, C(\epsilon) \rho\frac{\partial g(\epsilon)}{\partial \epsilon}\right|_{-\epsilon\gg 1}\, .
\end{equation}
Here, the lower limit lies in an energy interval some $\theta$ below the barrier top where the nonequilibrium distribution matches onto the Boltzmann distribution in the well. In this range $C(\epsilon)\approx 1$, while still $I(\epsilon)\approx I(\epsilon=0)=I(E=\Delta V)$.
To leading order in the friction strength we thus find the second main result, i.e.,
\begin{eqnarray}
\label{qmrate}
\Gamma_{\rm scl}=\frac{\sinh(\omega_0\hbar\beta/2)}{(\omega_0\hbar\beta/2)}\, |B|\ \Gamma_{\rm cl}
\end{eqnarray}
with the classical result specified in (\ref{classrate}). The first factor captures quantum effects (zero-point fluctuations) in the well distribution, while the second one describes the impact of finite barrier transmission close to the top. More details will be discussed in the next Section.

\section{Discussion}\label{discuss}
Of particular interest are the leading quantum corrections to the classical rate expression when either of the parameters $\rho$ or $\theta$ becomes small. For somewhat larger friction (but still sufficiently away from the turnover range) higher order corrections can also be accounted for along the lines described in \cite{rips_90}.

\subsection{Quantum effects}
To further analyze quantum effects in (\ref{qmrate}) we expand the prefactor ${\cal Y}=\Gamma_{\rm scl}/\Gamma_{\rm cl}={\cal Y}(\theta, \rho)$  in the inverse temperature $\theta$ as well as in the friction strength $\rho$. From (\ref{diffg}) one observes, however, that the limits $\rho\to 0$ and $\theta\to 0$ are not interchangeable: In the classical range and for energies below the barrier top the right hand side vanishes according to $T(\epsilon)/R(\epsilon)=\exp(\epsilon/\theta)\to 0$, while for any finite $\theta$ it diverges for $\rho\to 0$. This behavior reflects the fact that quantum mechanically  a steady state with a finite flux out of the well exists even in absence of dissipation.
Hence, we consider first for fixed $\rho$ quantum corrections to the classical high temperature limit $\theta\to 0$ and gain
\begin{eqnarray}
{\cal Y}\approx 1-\theta\,  b_1+\theta^2\, (b_2+\pi^2/6)+O(\theta^3/\rho)\, .
\end{eqnarray}
\begin{figure}
\vspace*{0.1cm}
\epsfig{file=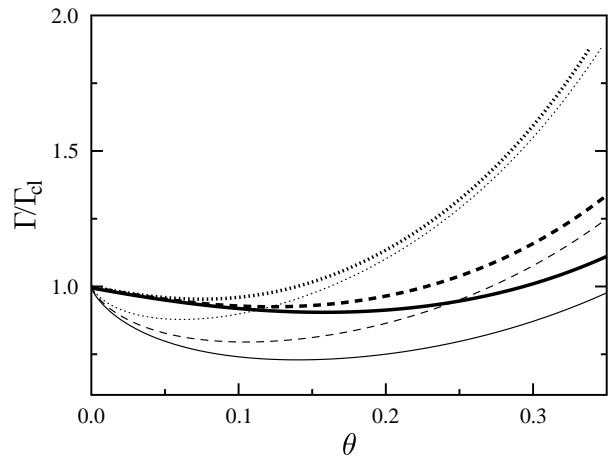, height=6cm}
\caption[]{Escape rate normalized to the classical rate (\ref{classrate}) versus the dimensionless inverse temperature $\theta=\hbar\omega_{\rm b}\beta$. Thin lines refer to the result from \cite{griff_89} where $C(E)=1$ [cf.~\ref{LEQM})], thick lines to the new expression (\ref{qmrate}) for $\gamma/\omega_0$
=0.0015 (solid), 0.005 (dashed), 0.01 (dotted)\label{fig2}.} \vspace*{-0.25cm}
\end{figure}
Here, the coefficient $b_1\simeq 1.04$ originates merely from the expansion of $B$ and describes the impact of finite barrier transmission, while $b_2\simeq 0.20$ is determined also by the well partition function. An inverse friction dependence appears in third order terms, which in turn means that the above expansion is only valid provided $\rho$ remains finite and that its range of validity shrinks with decreasing $\rho$.
Interestingly, there exists a temperature range with an upper bound given by $b_1/(b_2+\pi^2/6)\approx 0.56$, where quantum effects {\em reduce} the escape rate below its classical value and only for lower temperatures, i.e.\ larger $\theta$, does  the quantum result exceed the classical one (see fig.~\ref{fig2}).
This finding has already been predicted in \cite{griff_89,rips_90} with a quadratic temperature dependence in leading order though, which relates directly to the fact that {\em classical} transition matrix elements have been used in these previous works. Apparently, the impact of quantum fluctuations in (\ref{qmrate}) is enhanced and the suppression of the escape due to a finite reflection at the barrier top is diminished. It is only for lower temperatures that the quadratic dependence on the inverse temperature dominates and always leads to a rate increase compared to the classical result.

\begin{figure}
\vspace*{0.1cm}
\epsfig{file=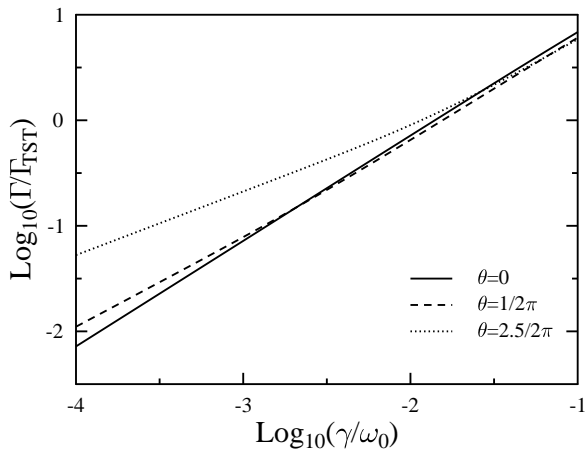, height=6cm}
\caption[]{\label{fig3}Escape rate normalized to the classical TST  rate ($\Gamma_{TST}=\omega_0/2\pi e^{-\beta\Delta V}$)  as a function of the dimensionless friction strength for various values of the dimensionless inverse temperature $\theta=\hbar\omega_{\rm b}\beta/2\pi$ ($\theta=0$ denotes the classical rate)} \vspace*{-0.25cm}
\end{figure}

The leading quantum correction for $\rho\to 0$ at fixed temperature can already be read off (\ref{diffg}).
Namely, the energy range below the barrier top where the flux solution matches onto the thermal equilibrium grows
for $\rho\to 0$ according to $\epsilon \sim \theta {\rm ln}(\rho)$. To leading order one then obtains in this domain
 $\partial_\epsilon g\propto -\exp(\epsilon)/\rho^\theta$, where the proportionality constant is determined by the boundary conditions and must be obtained from the full solution (\ref{solu1}). The leading order quantum correction to the rate prefactor scales thus like ${\cal Y}\propto \rho^{-\theta}$, which is non-analytical in $\rho$ \cite{rips_90}. The detailed calculation using large parameter expansions of the hypergeometric functions \cite{jones_01} gives
\begin{equation}\label{exprho}
{\cal Y}\approx\frac{\sinh(\omega_0\hbar\beta/2)}{(\omega_0\hbar\beta/2)}\,\frac{\Gamma(1-\theta)}{\Gamma(1+\theta)}\ \theta ^{2\theta}\, \rho^{-\theta}+O(\rho^{-\theta+1/2})\, .
\end{equation}
This behavior is illustrated in fig.~\ref{fig3}, where the rate becomes extremely sensitive to $\theta$ for weaker friction and lower temperatures.
It is interesting to note that for $\rho\rightarrow0$ and sufficiently high temperatures our result (\ref{exprho}) coincides with the result obtained previously in \cite{griff_89} since in this limit the matching happens for energies below the barrier top where $C=1$ in (\ref{diffg}).

\subsection{Higher order friction corrections}

As already mentioned at the end of Sec.~\ref{classical} the energy diffusive limit can also be described by a normal mode analysis in full configuration space. In \cite{rips_90} this approach has been applied to include quantum effects due to a finite barrier transmission, while transition matrix elements have been treated purely classical. Based on the above analysis the generalization of the latter ones to the semiclassical domain is straightforward. Of course, to leading order in the dissipation strength one recovers (\ref{qmrate}). Higher order corrections (still above turnover though) appear on the one hand through $\omega_{\rm b}\to \lambda_{\rm b}$ [see (\ref{grote})] in the factor $B$, thus capturing the impact of friction in the tunneling process. On the other hand, the full partition function (\ref{partgamma}) of the damped harmonic oscillator must be used in the well together with a factor describing the influence of the stable normal modes around the barrier top. This way, one gets
\begin{equation}
{\cal Y}=\frac{1}{\omega_0\hbar\beta Z_0}\, |B(\omega_b\to \lambda_{\rm b})|\, \prod_{n=1}^{\infty}\frac{\nu_n^2-\omega_{\rm b}^2}{\nu_n^2-\omega_{\rm b}^2+\nu_n \gamma}\, .
\end{equation}

\acknowledgements
Fruitful discussions with H. Grabert and E. Pollak are gratefully acknowledged. Financial support was provided by the German-Israeli Foundation through grant G-790-113.14/2003.

\appendix

\section{Moments of energy fluctuations}\label{appA}

In this Appendix, we specify how to  write the master equation (\ref{eq:master_d2}) in a way, which is amenable to a systematic semiclassical expansion.
Therefore we can use the $\hbar$-expansion of the delta function
\begin{eqnarray*}\label{eq:trucco1}
\delta(E-E_n-\hbar A_l(E_n))f(E_n)=[\delta(E-E_n)+\nonumber\\
\sum_{k=1}^{\infty}\left(\frac{\partial}{\partial E}\right)^k\frac{(-\hbar A_l(E_n))^k}{k!}\delta(E-E_n)]f(E_n)
\end{eqnarray*}
and rearrange the sum in (\ref{eq:master_d2}) in the following way,
\begin{eqnarray*}\label{eq:trucco2}
\sum_{n=0}^{N}\sum_{\stackrel{l=-n}{l\neq0}}^{N-n}f(n,l)=
\sum_{l=1}^{N}\left[\sum_{n=0}^{N}\left(f(n,l)+f(n,-l)\right)\right.\\
\left.-\sum_{n=0}^{l-1}f(n,-l)-\sum_{n=N-l+1}^{N}f(n,l)\right]\, .
\end{eqnarray*}
In (\ref{eq:master_d2}), there is  $f(n,l)=W_l(E_n)P_n$ and
\begin{eqnarray*}
W_{-l}(E_n)=\langle E_{n-l}|q|E_n\rangle=0& & \mbox{if } n\leq l-1\\
W_{l}(E_n)=\langle E_{n+l}|q|E_n\rangle=0& & \mbox{if } n\geq N-l+1
\end{eqnarray*}
Therefore the last two sums  in (\ref{eq:trucco2}) vanish.
Eqs.~ (\ref{eq:trucco1}) and~(\ref{eq:trucco2}) may be inserted in (\ref{eq:master_d2}) to give the final result given in (\ref{hbardiffus}).

\section{Coefficients for transition matrix elements}\label{appB}
Here we collect the coefficients appearing in the $\hbar$-expansion of $|Q_{qm}|^2$ in (\ref{eq:w_c}).
One has
\[
\tilde{A}_{\rm qm}=|Q_{\rm cl}|^2(R^2+1)+R(Q_{\rm cl}^2+Q_{\rm cl}^{*^2})
\]
and
\[
\tilde{B}_{\rm qm}= 2\tilde{N}^3\tilde{N}^{\prime}[|Q_{\rm cl}|^2(R^2+1)+R(Q_{\rm cl}^2+Q_{\rm cl}^{*^2})]
\]
as well as
\begin{eqnarray*}
\lefteqn{\tilde{C}_{\rm qm}=\frac{1}{2}(R^2+1){|Q_{\rm cl}|^2}^{\prime}+\frac{1}{2}R [(Q_{\rm cl}^2)^{\prime}+(Q_{\rm cl}^{*^2})^{\prime}]}\nonumber\\
&&+R|Q_{\rm cl}|^2 R^{\prime}+r^* r^{\prime} Q_{\rm cl}^{*^2}+r {r^*}^\prime Q_{\rm cl}^2\\
&&+(R^2+1)(Q_{\rm cl} K^*+Q_{\rm cl}^* K)+2 R(Q_{\rm cl}K+Q_{\rm cl}^*K^*)\, .
\end{eqnarray*}

\section{Expansion of the diffusion operator}\label{appC}

In this Appendix we will expand the moments of the energy fluctuations $ \langle\delta\rangle$ and $ \langle\delta^2\rangle$ in order to be able to write a semiclassical expression for ${\mathcal L}_E^{(\rm scl)}$.

In a first step we write $ \langle\delta\rangle$  with (\ref{bathdensity}),(\ref{eq:Wq}), (\ref{eq:epsilon}) and (\ref{eq:density}) as
\begin{eqnarray*}
\langle\delta\rangle=\frac{2 m\gamma\omega(E)}{\hbar}\int_{-\infty}^{\infty}d\delta^2\, \delta \, \overline{n}(\delta)\, |Q_{\rm qm}|^2.
\end{eqnarray*}
With the expansion of the Bose occupation function
\[
\overline{n}(\delta)=\frac{1}{{\rm e}^{\beta \delta}-1}\approx\frac{1}{\beta\delta}-\frac{1}{2}\, ,
\]
with (\ref{eq:w_c}) and the results of \ref{appB} we obtain the $\hbar-$expansion
\begin{eqnarray*}
\langle\delta\rangle=\frac{2\omega m\gamma}{\hbar}\int_{-\infty}^{\infty}d\delta\delta^2\left[ \frac{\tilde{N}^4\tilde{A}}{\beta\delta}-\frac{\tilde{N}^4\tilde{A}}{2}\right.\\
\left.+\frac{(\tilde{B}+\tilde{N}^4\tilde{C})}{\beta}\right].
\end{eqnarray*}
The integration leads to
\begin{eqnarray}
& &\langle\delta\rangle=\frac{\gamma\hbar^2\omega}{2\pi}\left\{\frac{-\tilde{N}^4}{2}(R^2+1)\omega^2I\right.\nonumber\\
&&\hspace{.4cm}+\left.\frac{1}{\beta}\left[(R^2+1)\tilde{N}^3\tilde{N}'\omega^2I+\frac{R^2+1}{4}
\tilde{N}^4(\omega^2I)'\right.\right.\nonumber\\
&&\hspace{3.9cm}\left.\left.+RR'\frac{\tilde{N}^4}{2}\omega^2I\right]\right\}\, ,
\label{delta1}
\end{eqnarray}
where $I$ denotes the action (\ref{periodicaction}).

In the same way one arrives at
\begin{eqnarray}
\langle\delta^2\rangle&=&\frac{\omega(E)m\gamma}{\hbar}\int_{-\infty}^{\infty}d\delta\delta^3
\frac{\tilde{N}^4\tilde{A}}{\beta\delta}\nonumber\\
&=&\frac{\gamma\hbar^2\omega}{2\pi\beta}(R^2+1)\frac{\tilde{N}^4}{4}\omega^2I\, .
\label{delta2}
\end{eqnarray}

Now we are able to combine (\ref{delta1}) and (\ref{delta2}) with (\ref{eq:L_scl}) to get the evolution operator in the energy diffusive regime (\ref{LEQM}).

\bibliographystyle{unsrt}
\bibliography{biblio}

\begin{thebibliography}{10}

\bibitem{benson_60}
S.~Benson.
\newblock {\em The Foundation of Chemical Kinetics}.
\newblock McGraw-Hill, New York, 1960.

\bibitem{haenggi_90}
P.~H{\"{a}}nggi, P.~Talkner, and M.~Borkovec.
\newblock {\em Rev. Mod. Phys.}, 62:251, 1990.

\bibitem{haenggi_93}
P.~H{\"a}nggi and G.R.~Fleming (eds.).
\newblock {\em Activated Barrier Crossing}.
\newblock World Scientific, New York, 1993.

\bibitem{pollak_05}
E.~Pollak and P.~Talkner.
\newblock {\em Chaos}, 15:026116, 2005.

\bibitem{vion_02}
D.~Vion, A.~Aassime, A.~Cottet, P.~Joyez, H.~Pothier, C.~Urbina, D.~Esteve, and
  M.~H. Devoret.
\newblock {\em Science}, 296:886, 2002.

\bibitem{claudon_04}
J.~Claudon, F.~Balestro, F.~W.~J. Hekking, and O.~Buisson.
\newblock {\em Phys. Rev. Lett.}, 93:187003, 2004.

\bibitem{martinis_02}
J.M. Martinis, S.~Nam, J.~Aumentado, and C.~Urbina.
\newblock {\em Phys. Rev. Lett.}, 89:117901, 2002.

\bibitem{siddiqui_04}
I.~Siddiqui, R.~Vijay, C.~M. Wilson, M.~Metcalfe, C.~Rigetti, L.~Frunzio, and
  M.~H. Devoret.
\newblock {\em Phys. Rev. Lett.}, 93:207002, 2004.

\bibitem{metcalfe_07}
M.~Metcalfe, E.~Boaknin, R.~Vijay, I~Siddiqi, L.~Frunzio, R.~J. Schoelkopf, and
  M.~H. Devoret.
\newblock {\em Phys. Rev. B}, 76:174516, 2007.

\bibitem{marthaler_06}
M.~Marthaler and M.~I. Dykman.
\newblock {\em Phys. Rev. A 73}, 73:042108, 2006.

\bibitem{dykman_90}
M.~I. Dykman and V.N. Smelyanski.
\newblock {\em Phys. Rev. A}, 41:3090, 1990.

\bibitem{kramers_40}
H.~A. Kramers.
\newblock {\em Physica}, 7:284, 1940.

\bibitem{caldeira_81}
A.~O. Caldeira and A.~J. Legett.
\newblock {\em Phys. Rev. Lett.}, 46:211, 1981.

\bibitem{grabert_84}
H.~Grabert and U.~Weiss.
\newblock {\em Phys. Rev. Lett.}, 53:1787, 1984.

\bibitem{leggett_92}
Yu. Kagan and A.J.~Leggett (eds.).
\newblock {\em Quantum Tunneling in Condensed Media}.
\newblock Elsevier, Amsterdam, 1992.

\bibitem{weiss_07}
U.~Weiss.
\newblock {\em Quantum Dissipative Systems}.
\newblock World Scientific, Singapore, 2007.

\bibitem{grabert_87}
H.~Grabert, P.~Olschowski, and U.~Weiss.
\newblock {\em Phys. Rev. B}, 36:1931, 1987.

\bibitem{ankerhold_07}
J.~Ankerhold.
\newblock {\em Quantum Tunneling in Complex Systems}.
\newblock Springer, Berlin, 2007.

\bibitem{rips_86}
I.~Rips and J.~Jortner.
\newblock {\em Phys. Rev. B}, 34:233, 1986.

\bibitem{dekker_88}
H.~Dekker.
\newblock {\em Phys. Rev. A}, 38:6351, 1988.

\bibitem{griff_89}
U.~Griff, H.~Grabert, P.~H{\"{a}}nggi, and P.S. Riseborough.
\newblock {\em Phys. Rev. B}, 40:7295, 1989.

\bibitem{rips_90}
I.~Rips and E.~Pollak.
\newblock {\em Phys. Rev.A}, 41:5366, 1990.

\bibitem{coffey_07}
W.~T. Coffey, Yu.~P. Kalmykov, S.~V. Titov, and B.~P. Mulligan.
\newblock {\em PCCP}, 9:33, 2007.

\bibitem{larkin_86}
A.~I. Larkin and Yu.~N. Ovchinnikov.
\newblock {\em Sov. Phys. JETP}, 64:185, 1986.

\bibitem{caldeira_83}
A.~O. Caldeira and A.~J. Leggett.
\newblock {\em Physica A}, 121:587, 1983.

\bibitem{zwanzig_73}
R.~Zwanzig.
\newblock {\em J. Stat. Phys.}, 9:215, 1973.

\bibitem{pollak_86_A}
E.~Pollak.
\newblock {\em J. Chem. Phys.}, 85:865, 1986.

\bibitem{pollak_89_A}
E.~Pollak, H.~Grabert, and P.~H{\"{a}}nggi.
\newblock {\em J. Chem. Phys.}, 91:4073, 1989.

\bibitem{grote_80}
R.~F. Grote and J.~T. Hynes.
\newblock {\em J. Chem. Phys.}, 73:2715, 1980.

\bibitem{karrlein_97}
R.~Karrlein and H.~Grabert.
\newblock {\em J. Chem. Phys.}, 108:4972, 1997.

\bibitem{breuer_02}
H.-P. Breuer and F.~Petruccione.
\newblock {\em The Theory of Open Quantum Systems}.
\newblock Oxford, 2002.

\bibitem{dunham_32}
J.~L. Dunham.
\newblock {\em Phys. Rev.}, 41:713, 1932.

\bibitem{more_91}
R.~M. More and K.~H. Warren.
\newblock {\em Ann. Phys.}, 207:282, 1991.

\bibitem{atabek_91}
O.~Atabek, R.~Lefebvre, M.~Garcia Sucre, J.~Gomez-Llorente, and H.~Taylor.
\newblock {\em Int. J. Quant. Chem.}, 40:211, 1991.

\bibitem{melnikov_85}
V.I. Melnikov.
\newblock {\em Physica}, A130:606, 1985.

\bibitem{jones_01}
D.~S. Jones.
\newblock {\em Math. Meth. Appl. Sci.}, 24:369, 2001.

\end{thebibliography}

\end{document}